\documentclass{PoS}
\usepackage{amssymb,fontenc,times,mathptmx,graphicx,epsfig,xspace}

\newcommand {\ie}{\mbox{i.e. }\xspace}     
\newcommand {\etc}{\mbox{etc. }\xspace}     

\newcommand{\fbinv} {\mbox{\ensuremath{\mathrm{\; fb}^{-1}}}\xspace}

\newcommand{\hilumi} {\ensuremath{{\cal{L}}=10^{34}\; \mathrm{cm^{-2}\,s^{-1}}}\xspace}

\newcommand{\bbbar}{\ensuremath{{b\overline{b}}}\xspace}

\newcommand{\ttbar}{\ensuremath{{t\overline{t}}}\xspace}

\newcommand{\PT}{\ensuremath{p_{\mathrm{T}}}\xspace}

\newcommand{\ET}{\ensuremath{E_{\mathrm{T}}}\xspace}

\newcommand{\MET}{\ensuremath{E_{\mathrm{T}}^{\mathrm{miss}}}\xspace}

\def\ga{\mathrel{\rlap{\raise.6ex\hbox{$>$}}{\lower.6ex\hbox{$\sim$}}}}
\def\la{\mathrel{\rlap{\raise.6ex\hbox{$<$}}{\lower.6ex\hbox{$\sim$}}}}
\newcommand{\rpv}{\ensuremath{\rlap{\kern.2em/}R}\xspace}

\newcommand{\ra}{\rightarrow}
\def\signal{$H\rightarrow W^{+}W^{-} \rightarrow l^{+}\nu l^{-}\bar{\nu}$~}

\title{Top Background to SM Higgs Searches in the $W^{+}W^{-} \ra \ell \nu \ell \bar{\nu}$ 
  Decay Mode at CMS}
\author{
  G.Davatz \\ Institute for Particle Physics, ETH Zurig, Switzerland 
}
\author{
  A.S. Giolo-Nicollerat \\ CERN, Geneva, Switzerland 
}
\author{
  \speaker{M.Zanetti} \\ Dipartimento di Fisica {\em ``Galileo Galilei''} 
  and INFN, Padova, Italy 

}

\abstract{The top quark and its properties within and beyond the Standard Model
  will be extensively studied at the incoming Large Hadron Collider.
  Nonetheless the top quark will play the role of the main background for most
  of the Higgs and new physics searches. In this paper the top as a background to
  \signal Higgs discovery channel will be studied. The current status of the Monte Carlo tools
  for \ttbar and single top simulation will be presented. Finally the problem
  on how to evaluate the top background from the data will be addressed and the
  related systematics will be discussed.
} 
\FullConference{Top2006, Coimbra, Portugal}
\ShortTitle{Top background to \signal at CMS}

\begin{document}

\section{Introduction}

At the Large Hadron Collider (LHC), 
the production cross section for processes involving the
top quark, either in the singly and the doubly resonant mode,
is foreseen to exceed 1 nb (respectively $\sim{0.3}$ nb for single top 
and $\sim{0.8}$ nb for \ttbar). 
At the designed instantaneous luminosity of \hilumi,
averagely more than 10 events containing top per second are expected. 
In this sense the LHC will be a top factory allowing to study
with no statistical limitation the properties of the most peculiar particle 
discovered so far. Moreover, providing copiously most of the detectors'
observables like (b-)jets, leptons, missing transverse energy (\MET) \etc, 
top events will be
a candle for the comprehension of various experimental systematics and 
detector performances. Nevertheless, at the same time, the top quark will play 
the role of background for most of the Higgs and new physics signatures.

%
%
Studying the top quark as a background instead as a signal is 
clearly a complete different issue that needs different 
experimental and theoretical approaches with consequent 
specific systematics uncertainties. Moreover, 
the phase space where the top plays the role of the background 
is usually narrow and contains only the tails of the events' distributions. \\
In this paper the top background to the \signal Higgs discovery channel
will be studied in details for the case of the CMS detector. \\
In the first section, a brief overview of the characteristics of 
the signal and the top background will be given. 
The following part will focus on the simulation of the top events;
the generation of \ttbar events will be studied by comparing different 
Monte Carlo programs, then the inclusion of singly
resonant top production at Next to Leading Order (NLO) will be discussed. \\ 
The LHC energy regime has never been probed so far, thus any
sound analysis must rely on measured data to the maximum extent. 
In the second part, the problem on how to normalize the top background using
data will be addressed. Finally the experimental uncertainties
coming from different normalization strategies will be estimated
using a full CMS simulation.

\section{The $H\ra WW\ra \ell \nu \ell \nu$ and the top background}
\label{signalDescription}

\begin{figure}[ht]
  \begin{center}
    \includegraphics[width=0.37\textwidth]{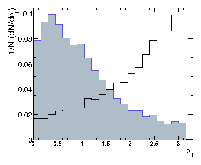} 
    \hspace{0.5 cm}
    \includegraphics[width=0.3\textwidth, angle=90]{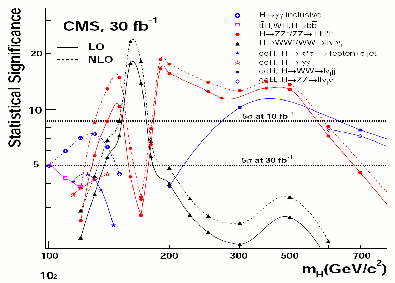} 
  \end{center}
  \caption{Left: opening angle between the leptons ($\rm \phi_{\ell\ell}$) for the
    signal (blue) and the W-pairs background (black). 
    Right: statistical significance foreseen with 30 \fbinv at CMS for 
    various Higgs channels as a function of  m$_{\mathrm{H}}$.
 } 
\label{signalPlots}
\end{figure}

At the LHC, the \signal Higgs decay mode is considered the most favorable
one for a Higgs mass ranging between $150-170$ GeV~\cite{dittmar}~\cite{ASGD}. 
This can be argued from the right plot in Figure \ref{signalPlots}, showing 
the statistical significance foreseen with 30 \fbinv at CMS for the
various Higgs channels as a function of  m$_{\mathrm{H}}$. \\
The signal is characterized by two isolated, high \PT and nearby leptons with
small invariant mass, an high value of \MET and no reconstructed jet
in the central part of the detector. The main backgrounds are the 
non-resonant $W$-pair, the Drell Yan, the \ttbar and the single top in 
the $Wt$ mode\footnote{One may think to consider the complete gauge 
  invariant process $W^+bW^-\bar{b} \ra l^{+}\nu  b l^{-}\bar{\nu} \bar{b}$
  instead of keeping separate \ttbar and $tWb$. The reason why this 
  is not done will be given in section~\ref{tWDescription}
}. \\
The variable that allows to discriminate the irreducible background 
from continuous $W$-pair production is the opening angle between the leptons.
Due to scalar nature of the Higgs, the $W$ bosons are produced mainly with
opposite helicity, then, because of the V-A structure of the $W$s coupling to leptons,
the latters are produced nearby in the space. In the left plot
of Figure \ref{signalPlots} the opening angle between the leptons are shown
for the signal and the $q\bar{q} \ra W^+W^-$ process.
Since the angular distributions of the leptons are crucial variables, the
top polarization should be taken into account throughout its whole decay chain
by the Monte Carlo generators. This issue will be addressed in \ref{spinCorrDescirption}

The most powerful cut to reduce the top events is the jet veto which requires
not to have reconstructed jets above a certain \ET threshold in the central part of 
the detector. Dealing with low \ET jets in the LHC environment is a delicate issue 
either from the experimental and from the theoretical point of view. 
More details about this selection will be given in \ref{signalRegion}. 
When the problem of the NLO description of $Wt$ will addressed, it will also be remarked 
how the application of a jet veto helps in consistently separate this process 
from \ttbar.

\begin{table}[h!]
\begin{center}
\begin{tabular}{|l|l|c|c|c|}
\hline
& & $\rm H\to WW$ ($\rm m_H=$~165~GeV) &  \ttbar & $\rm tWb$ \\
\hline
&$\rm \sigma \times BR(e,\mu,\tau)$ [fb] &2360 & 86200 & 3400\\
\hline
1) & Trigger & 1390 (59\%) &  57380 (67\%) & 2320 (68\%)  \\
\hline
2) & lepton ID & 393 (28\%) &15700 (27\%) & 676 (29\%)  \\
\hline
3) & $\rm E_t^{miss}>$~50~GeV & 274 (70\%) &9332 (59\%) & 391 (58\%) \\
\hline
4) & $\rm \phi_{\ell\ell}<$~45 & 158 (58\%) &1649 (18\%) & 65 (17\%)\\
\hline
5) & 12~GeV~$<m_{\ell\ell}<$~40~GeV & 119 (75\%) & 661 (40\%) &
28 (43\%)\\
\hline
6) & 30~GeV$\rm <p_{t}^{\ell\, max}<$55~GeV & 88 (74\%) &304 (46\%)
& 13 (46\%)\\
\hline
7) & $\rm p_{t}^{\ell\, min}>$25~GeV & 75 (85\%) & 220 (73\%) & 9.2 (71\%)\\
\hline
8) & Jet veto & 46 (61\%) &  9.8 (4.5\%) & 1.4 (15\%)\\
\hline
\end{tabular}
\end{center}
\caption{The expected number of events for a luminosity of 1~fb$^{-1}$
for the signal with a Higgs mass of 165~GeV and the  \ttbar and
$\rm tWb$ background. The relative efficiency with respect to
the previous cut is given inside the brackets in percent.}
\label{signal_selections}
\end{table}

In Table~\ref{signal_selections} the complete list of selections used  
together with the 
corresponding number of events expected for 1~fb$^{-1}$ for the fully simulated 
signal (for a Higgs mass of 165 GeV), \ttbar and $Wt$ are summarized.
The rejection for \ttbar is $\rm {\cal O}(10^{-4})$, which sets a challenge for
the needed precision of the Monte Carlo calculations. 
Moreover the presence of two neutrinos in the signal final state 
does not allow the reconstruction of a narrow invariant mass peak; 
the discovery has then 
to rely on an excess in the expected number of background events.
It is thus necessary to identify a phase space region where reliably
control the contribution of the different backgrounds.
In section \ref{normalizationDescription},
two procedure for the \ttbar normalization are proposed and discussed.

\section{\ttbar background generation}

In this section the generation of top-pair process 
($\rm pp\ra t\bar{t} \ra WbW\bar{b}\ra
\ell\nu\ell\nu b\bar{b}$, with $\rm \ell=e$, $\rm \mu$ and $\rm \tau$) 
will be discussed by comparing four different Monte Carlo (MC) generators.
Three points will be addressed: first we will estimate
how much the Born level description differs from the NLO
one in the phase space relevant for the Higgs search.
Then we will determine whether different showering models 
cause differences in the relevant kinematics distributions.
Finally how much the inclusion of the spin correlation between the 
top quarks affects the leptons' angular distribution will be
estimated. 
All the following studies have been done at parton level,
without exploiting a full detector simulation.

\subsection{NLO effects on \ttbar simulation}

To estimate the effect of an accurate inclusion of NLO matrix elements, 
HERWIG~6.508~\cite{herwig} (LO, parton shower Monte Carlo) 
and MC@NLO~2.31~\cite{mcatnlo} (NLO, parton shower Monte Carlo) were compared.
The spin correlations between the $\rm t$ and $\rm \bar{t}$ are not considered 
in MC@NLO. HERWIG events were therefore consistently simulated without such correlation. 
As the same showering model is used, the difference between the two simulations 
should be mostly due to the additional NLO matrix elements in MC@NLO. 

\begin{figure}[h!]
\begin{center}
\includegraphics[width=1\textwidth]{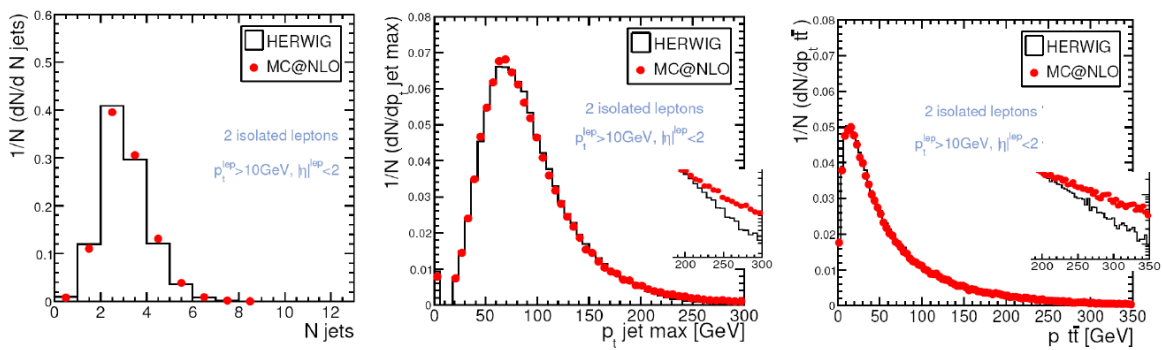}
\end{center}
\caption{Left: jet multiplicity. Center: leading jet \ET. Right: transvers momentum
  of the \ttbar system. The little windows in the center and right plots show the deviation
  at high \PT in logarithmic scale.
} 
\label{nloShapes}
\end{figure}

In Figure~\ref{nloShapes}, from left to right, the number of jets, the
\ET of the leading jet and the \PT of the \ttbar system are shown\footnote{
  this variable gives an estimation of the spectrum of the additional
  jets to the hard process}.
MC@NLO produces in addition to the hard process up to one hard jet 
whose spectrum is accurate at NLO. Most of the jet activity 
in the events is however dominated by the two b-quarks
from the two top quarks decay. None of the three distributions 
shows indeed relevant differences between the NLO and LO.  
Typical NLO effects can be noticed in the high part of
the spectrum either in the leading jet \ET distribution and in the
\ttbar system transverse momentum: the parton shower is in fact
known to describe correctly only the soft part of the extra
jet activity. 
Nevertheless the region relevant for the $\rm  H\to WW\to \ell \nu \ell \nu$ 
signal selection is the very low $\rm p_t$ region, 
where HERWIG and MC@NLO agree very well. 
In addition, the shapes of all the other cut
variables are very similar in MC@NLO and HERWIG without spin
correlations. \\
When comparing the relative efficiencies of the different cuts, 
the two Monte Carlos differ essentially only for the jet veto cut, the difference 
in the efficiency being  ${\cal O}(10\%)$~\cite{ttCMSPaper}. 
Since the region where NLO makes a difference is at very high $\rm p_t$, 
whereas the bulk of the selected events is in the low $\rm p_t$ region, 
it is safe to conclude that NLO effect can be simply included
by rescaling the cross section by an inclusive factor.

\subsection{Effect of showering models, differences between HERWIG and PYTHIA}

In the following, how different showering models influences the variable shapes and
selection efficiencies will be studied. For this, PYTHIA 6.325~\cite{pythia}, 
based on the Lund hadronization model, was compared with
HERWIG based on the cluster model for hadronization\footnote{PYTHIA 
  does not take into account the spin correlations between the top and the anti-top ,
  then in order to consistently study only the differences caused by
  the showering model, HERWIG with disabled spin correlations
  has been used
}. 
For both simulations, default scales were chosen. 

\begin{figure}[h!]
\begin{center}
\includegraphics[width=1\textwidth]{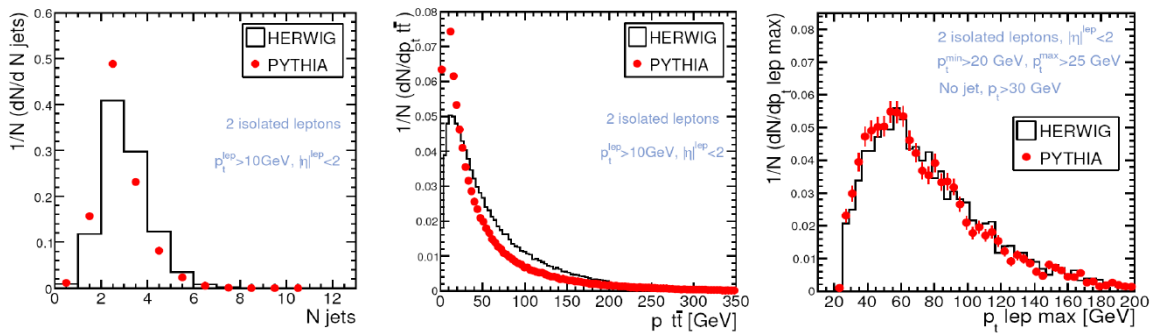}
\end{center}
\caption{Left: jet multiplicity. Center: \ttbar system \PT. 
  Right: \PT of the most energetic lepton after the jet veto }
\label{showerShapes}
\end{figure}

The left and central plots of Figure~\ref{showerShapes} shows
respectively the number of jets and the \PT of the \ttbar system.
The two showering models differ remarkably in both distributions, 
the Lund model predicting less and softer extra jets.
The effect of these discrepancies in the jet veto efficiencies is sizable, 
\ie about 20\%. \\
However, the shapes of the relevant leptons' kinematics distributions 
in the signal region are very similar for the two Monte Carlo's.
This can be seen from the right plot of of Figure~\ref{showerShapes}
showing the \PT spectrum of the most energetic lepton after applying 
the jet veto.

\subsection{Effect of the spin correlations}
\label{spinCorrDescirption}

As explained in~\ref{signalDescription}, the variable that 
characterizes more the signal is the opening angle between the two leptons.
This observable, as well as the mass of the dilepton system and the \MET, 
is sensitive to the spin correlations between the particles
involved in the process. In order to point out the effect of the
inclusion of the top polarization in the \ttbar decay chain on the 
leptons' angular distribution, PYTHIA has been compared with TopREX~\cite{TopREX}.
While the former does not consider the top spin along its decay, the latter
is a matrix element based Monte Carlo describing exactly 2$\ra$6 processes
with LO precision.

\begin{figure}[h!]
\begin{center}
\includegraphics[scale=0.7]{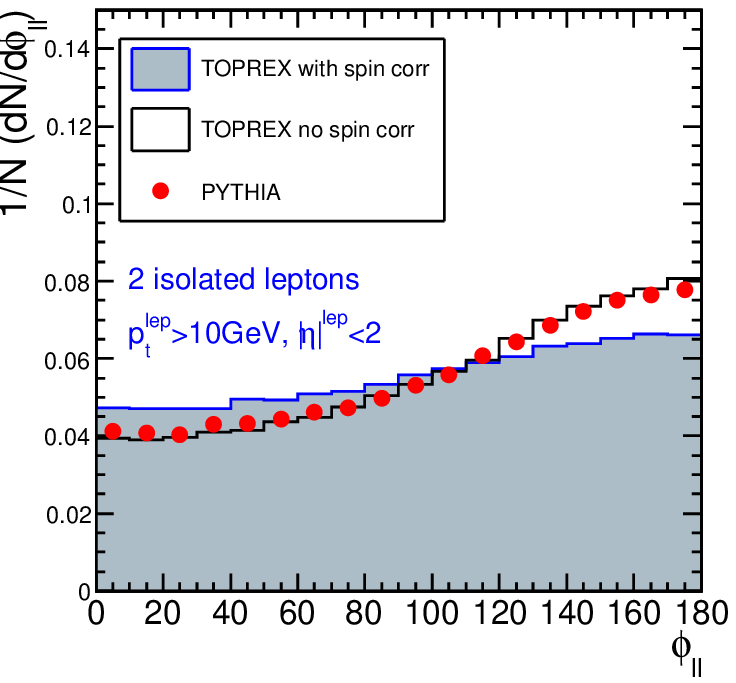}
\includegraphics[scale=0.7]{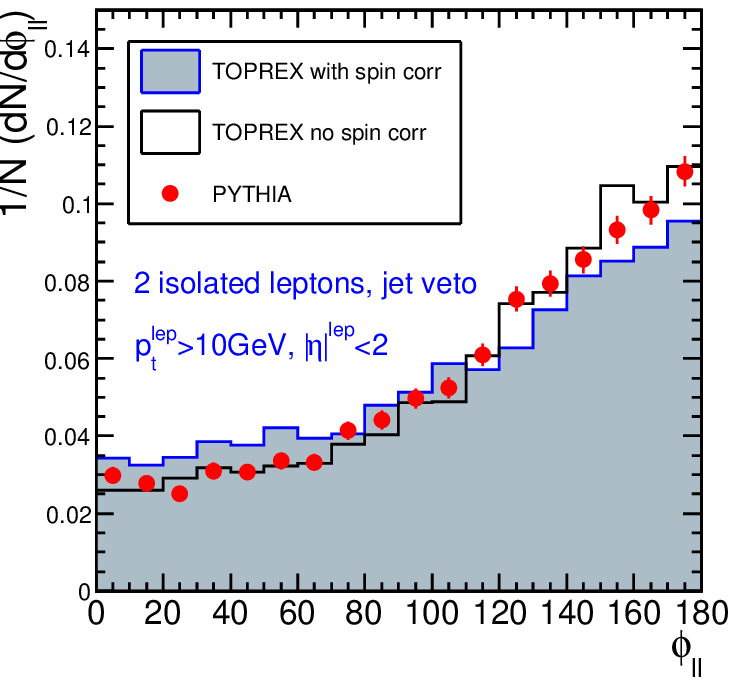}
\end{center}
\caption{$\rm \phi_{\ell\ell}$ distribution for TopREX events with and without spin correlations is shown, as well as PYTHIA. On the left, only very basic cuts are applied, whereas on the right a jet veto is applied in addition. The region important for the Higgs signal search is the low $\rm \phi_{\ell\ell}$ region.} 
\label{spinDPhiShapes}
\end{figure}

In the left plot of Figure~\ref{spinDPhiShapes} showing the angle $\rm
\phi_{\ell\ell}$ between the leptons a non negligible difference
between the two Monte Carlo's prediction can be seen. 
Quantitatively, the variation of the efficiency for the $\phi_{\ell\ell}$ cut
between the PYTHIA and TopREX is $\sim{10}\%$.
As it can be seen from the right plot of Figure~\ref{spinDPhiShapes}, 
in the signal region, \ie after applying the jet veto, the difference
between the two distributions, even if still present, tends to flat.


\section{Singly resonant top background generation}
\label{tWDescription}

\begin{figure}[htbp]
\begin{center}
\includegraphics[scale=0.8]{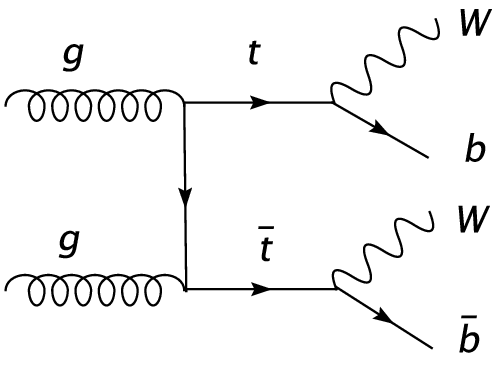} \hspace*{2cm}
\includegraphics[scale=0.8]{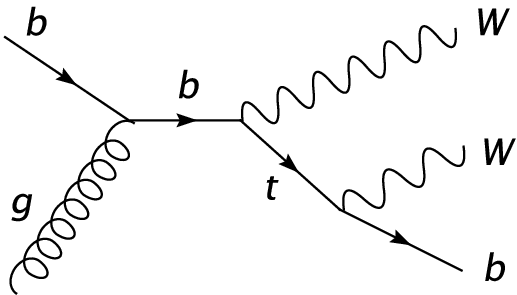}
\end{center}
\caption{Examples of Feynman graphs for double (left) and single (right)
top production}
\label{singleTopFeyn}
\end{figure}

Although the $Wt$ mode of the single top production has an estimated cross section
one order of magnitude times smaller than the \ttbar production, 
the application of the jet veto enhances the contribution of the former 
with respect to the latter's.
Considering also that the single top process has never been measured at the
Tevatron, it is important to pursuit a NLO description of this process
that allows a precise estimation of its features and its total cross section.
In principle one could think to use the singly and doubly resonant processes
together \ie to consider $pp\to WbWb\to \ell\nu\ell\nu bb$
which is naturally gauge invariant and describes correctly all the
interference terms. Nevertheless the NLO corrections are not available
for such a process and in particular it is not known how to deal with
the arising large logarithms of the form $\log ((m_t+m_W)/m_b)$.
In is therefore preferable to view the singly resonant process as one in which a
$\rm b$ quark is probed directly inside the proton~\cite{maltoniCoimbra} 
(right diagram in Figure~\ref{singleTopFeyn}). 

The diagrams contributing to the NLO description of $Wt$ includes the 
LO doubly resonant \ttbar (left diagram in Figure~\ref{singleTopFeyn}).
A solution to that has been provided by Campbell and Tramontano in Ref.~\cite{campbell} 
where they suggest to define a specific $Wt$ final state by imposing a veto
on the presence of an extra b quark. In this schema, also called ''b-PDF approach'',
the \PT threshold for the spectator b-quark (coming from an initial gluon splitting)
is set at the same value of the factorization scale used for the PDF. 

This approach perfectly fits with the case of the $Wt$ background for 
the \signal searches where a global jet veto is applied. 
In Ref.~\cite{ttCMSPaper} the efficiencies for $Wt$ events 
for the signal leptons' selections have been compared between 
LO plus parton shower ($ \rm TopREX + PYTHIA$)
and NLO (MCFM~\cite{mcfm}) descriptions, showing an overall agreement.
After matching the veto threshold on the b-quark with the threshold for the jet
veto, it is possible to consistently use the overall normalization 
provided by the NLO calculation for the cross section 
in the signal phase space area. The ratio between the LO and NLO cross section
goes from 1.4 before the application of the cuts, to 0.7 after the selection.
This is consistent with the fact that NLO calculations enhances the jet activity
thus reducing the jet veto efficiency.

\section{$\mathbf{t \bar{t}}$ normalization from data}
\label{normalizationDescription}

The commonly used method to normalize a given background from the data 
consists on selecting a signal-free 
phase space region (control region) where a given background process is enhanced. 
The contribution of that background in the signal region is then extrapolated
from the measured amount of events in the control region. 
This procedure relies on the relation:
\begin{equation} \label{normalization_relation}
  N_{signal\_reg}=\frac{N_{signal\_reg}^{Monte Carlo}}{N_{control\_reg}^{Monte Carlo}} N_{control\_reg}
  =\frac{\sigma_{signal\_reg}\cdot\epsilon_{signal\_reg}}{\sigma_{control\_reg}\cdot\epsilon_{control\_reg}} N_{control\_reg}
\end{equation}
where $N_{signal\_reg}^{Monte Carlo}$ and $N_{control\_reg}^{Monte Carlo}$ are the numbers 
of events predicted by the Monte Carlo simulation in the signal and control region.
Each of this two numbers can be expressed as a product of the theoretical cross
section in that phase space area, $\sigma_{signal\_reg,control\_reg}$, 
and the experimental efficiency of reconstructing
events in the same region, $\epsilon_{signal\_reg,control\_reg}$\footnote{
  The experimental uncertainties could modify the boundaries 
  defining the phase space where the cross section is calculated theoretically. 
  This is the case in particular when the selections involve jets. 
  The ``$\epsilon$'' terms in relation
  (\ref{normalization_relation}) are assumed to account also for this effect.
}.
This will allow to better point out the different sources of systematic uncertainties. 
In particular the theoretical predictions enter the procedure only via 
the ratio $\sigma_{signal\_reg}/\sigma_{control\_reg}$, leading to a much smaller scale dependency
and thus to smaller theoretical uncertainties. \\
The theoretical issues concerning the \ttbar normalization have been deeply studied 
in \cite{Kauer:2004fg}, following the work done in the 2003 Les Houches Workshop.
The primary goal here is to provide a reliable description of the experimental aspects,
specifically the ones related to the CMS detector. For this study a full detector simulation 
has then been exploited. \\
The main requirement from the experimental side on the choice of the control 
region is to limit as much as possible the error due to the ``$\epsilon$'' terms
in relation (\ref{normalization_relation}). This implies to use similar
selections as for the signal region. Moreover the contamination from
other physical and instrumental backgrounds should be negligible. \\
In the following the signal phase space area and two possible control regions 
for \ttbar normalization are described, focusing on the related experimental
issues. Each of these phase space regions (either the signal and the control ones)
is defined by the selections on the leptons listed in Table~\ref{signal_selections} (items 1-8).

\subsection{Signal region}
\label{signalRegion}

We already stated before that the signal region is defined by requiring 
not to have reconstructed jet above a certain \ET threshold in the central part
of the detector. Clearly the lower the \ET threshold, the higher
is the rejection for \ttbar background events. 
At CMS the definition of a jet at low \ET is experimentally problematic. 
This because the 4 Tesla magnetic field of the CMS solenoid 
spread the jet constituents in the transverse plane and  
prevents the low momentum charged tracks 
produced during the fragmentation even to reach the calorimeters. 
Moreover the LHC, in addition to the products of the hard scattering, 
produces thousands of charged and neutral particles some of which
may have rather high \PT, thus enhancing the jets fake rate. \\
In order to avoid a high rate of fake jets at low \ET, the tracking 
measurements are exploited in the jet definition. For jets between 15 and 20 GeV
the sum of the \PT of the tracks belonging to the jets (\ie the tracks 
coming from the primary vertex which stand within the jet cone) is 
required to be at least the 20\% of the jet \ET~\cite{ASGD}. 

\begin{figure}[h!]
\begin{center}
\includegraphics[width=0.8\textwidth]{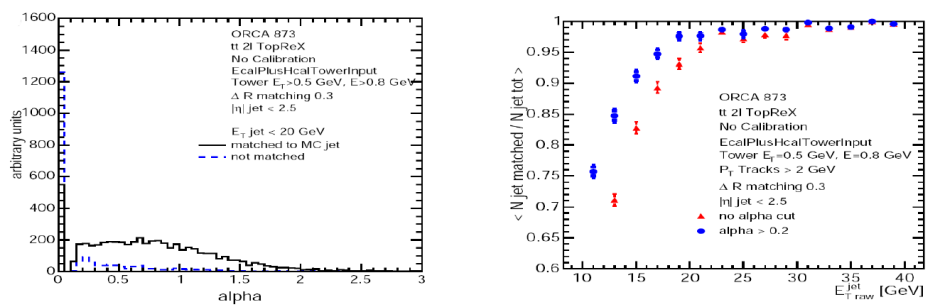}
\end{center}
\caption{Left: $\alpha$ distribution for matched and un-matched jets. 
  Right: fraction of matched jets as a function of the reconstructed
  jet \ET
}
\label{alphaHistos}
\end{figure}

As it can be seen from the right plot in Figure~\ref{alphaHistos}, the
fake jets (\ie not matched with any jet at parton level) rate below 20 GeV
decreases remarkably. The distribution of $\alpha=\Sigma\PT(tracks)/E_T(jet)$
from matched and un-matched jets is shown in the left plot of Figure~\ref{alphaHistos}.

\subsection{b-tagging jets based \ttbar control region}

The request for two b-tagged jets is the most natural for the
definition of a control region for the \ttbar background. 
In this study, the algorithm used to discriminate whether a jet is originated from
a $b$ quark is based on the impact parameters of charged particle tracks
associated to the jet~\cite{BTagging:2005an}.
The parameter 
that characterizes the efficiency and the mistagging rate of the 
algorithm is the impact parameter significance ($\sigma_{IP}$) 
of a minimum number of tracks associated to the jet. 
In this study a jet is tagged as a $b$-jet if its measured $E_t$ is greater
then $20~GeV$ and if there are at least 2 tracks whose 
$\sigma_{IP}$ is greater than 2.
In this case the double b-tagging efficiency is ${\cal O}(30\%)$ 
while the mistagging rate is ${\cal O}(3\%)$. 
Table ~\ref{doublebTag_table} 
summarizes the number of events expected for 10~fb$^{-1}$ in the
control region for \ttbar, $ Wt$ and the signal in the case of 
$\rm 2\mu$, $\rm 2e$ and $\rm e\mu$ final states.

Not all the processes with $2\ell+2b+E_t^{miss}$ as final state have been fully simulated
for this analysis, nevertheless general considerations and fast Monte Carlo
level checks can lead to exclude other relevant sources of backgrounds.  \\
The more natural concurrent process is the not resonant 
$W^{+}W^{-}\ra2\ell+b\bar{b}$ which is anyway
suppressed with respect to \ttbar. Its cross section is indeed
expected to be smaller than $1~pb$.   
Assuming the same efficiency for the kinematic selections as for the $W^{+}W^{-}\ra2\ell$,
i.e. ${\cal O}(10^{-3})$, less than $10$ events are aspected for $10~fb^{-1}$ in the control 
region even without folding the double-b tagging efficiency.  
In the case of same flavor leptons in the final state, $\gamma^{*}/Z^{*}\ra 2\ell + b\bar{b}$ 
(the vector boson mass being away from the $Z$ peek, i.e. $m_{\ell\ell}<40~ GeV$) 
could also contribute as an instrumental background, when an high value of $E_t^{miss}$ is
provided by the not full hermeticity of the detector and/or due the finite resolution
of the calorimeters. 
Anyway for a fully simulated sample of $\gamma^{*}/Z^{*}\ra 2\ell + 2b$ with jets' $E_t$ 
greater than $20$ GeV, the fraction of events with $E_t^{miss}>50~GeV$ 
(the actual cut applied for the signal selection) is  ${\cal O}(10^{-2})$.
Applying the same kinematic selections, but the $E_t^{miss}$ cut on a
$pp\ra \gamma^{*}/Z^{*}\ra 2\ell + b\bar{b}$ sample generated with MadGraph Monte Carlo
\cite{Maltoni:2002qb},
$200$ events are expected for $10~fb^{-1}$, which reduce to a negligible quantity if
the rejection due to a realistic $E_t^{miss}$ selection is included\footnote{
  In the $\gamma^{*}/Z^{*}\ra 2\ell + 2b$ fully simulated sample (the only one available) 
  the \bbbar pair
  comes only from a gluon splitting, the main mechanism of $\gamma^{*}/Z^{*}+ 2b$
  not being included. That is the reason why the selection cuts have been applied
  at parton level on a MadGraph sample}.

\subsection{Two high $E_t$ jets based \ttbar control region}

In order to avoid the systematics due to the b-tagging algorithm 
it is worth to have alternative methods to estimate  
the \ttbar background from data.
Each of the two $b$'s in the \ttbar final state come from a $175~GeV$ 
central object; their $E_t$ spectra are then rather hard. An alternative
method to define a \ttbar control region is thus simply to require, 
in addition to the signal kinematic cuts listed in Table~\ref{signal_selections}, 
two hard jets in the detector.\\
In order to avoid the contamination from Drell Yan which in the case of
$2\ell+ 2j$ final state has a much higher cross section then the $2\ell+ 2b$ one,
only $e\mu$ final state has been considered. \\ 
The thresholds on the jets' transverse energy that maximize the signal (\ttbar) 
over the background (Wt$+$signal) ratio and minimize the statistical error 
have been found to be $50$ and $30$ GeV. The number of events expected events
for $\rm 10~fb^{-1}$ for \ttbar, Wt and the signal are summarized 
in Table~\ref{doublebTag_table}.

A background process not considered in the full simulation analysis is
$W^{+}W^{-}\ra\mu\nu_{\mu}+e\nu_{e}+2j$. The cross section, after geometrical acceptance cuts,
is  $0.4~pb$, whereas the signal selection cuts efficiency
resulted to be smaller than $5\cdot 10^{-4}$ (with a statistical error of
$\sim{8}\%$). The contribution of this background can then be assumed to
be at maximum of the order as the signal. \\
In the case one jet is misidentified as an electron, $W^{\pm}\ra\mu\nu_{\mu}+3j$,
could be a source of background too.
At CMS, the probability of electron misidentification is 
estimated to be  ${\cal O}(10^{-4})$\footnote{
  The muon misidentification rate is at least one order of magnitude
  smaller}. 
Given its cross section, calculated to be $\sim{200}$ pb after the geometrical  
acceptance cuts, the latter rejection factor together
with the kinematic selection efficiency -estimated again from a generator level
study to be ${\cal O}(10^{-4})$- lead to neglect this process as a source 
of contamination of the \ttbar control region.

\begin{table}
  \begin{center}
    \begin{tabular} {|c|c|c|c|c|c|c|c|c|c|}
      \hline
      & \multicolumn{3}{c|}{``b-tagging'' control region} 
      & \multicolumn{3}{c|}{``hard jets'' control region} 
      & \multicolumn{3}{c|}{Signal region} \\
      \cline{2-10}
      & $2\mu$ & $2e$ & $e\mu$ & $2\mu$ & $2e$ & $e\mu$& $2\mu$ & $2e$ & $e\mu$ \\
      \hline
      \ttbar & 194 & 107 & 245 & - & - & 411 & 33 & 22 & 44\\
      \hline
      $Wt$  & 1 & $\rm <1$ & 2 & - & - & 6 & 5 & 3 & 6\\
      \hline
      $Signal~(m_H=165)$  & $\rm <1$ & $\rm <1$ & 1 & - & - & 11 & 156 & 89 & 214\\
      \hline
    \end{tabular}
  \end{center}
  \caption{Number of events of \ttbar, Signal and $Wt$ expected for $10~fb^{-1}$ 
    in the two control regions described above and in the signal region.
    Results are shown for $2\mu$, $2e$, $e\mu$ final states.
    \label{doublebTag_table}}
\end{table}

\section{\ttbar normalization procedure uncertainties}

Our proposed procedures to estimate the number of \ttbar events 
in the signal phase space region exploits relation (\ref{normalization_relation}).
In order to compute the systematic uncertainties on the final result we 
consider separately those related to each term present in the formula.


\subsection* {\bf Theoretical uncertainty.} 

  Taking the ratio of the
  $t\bar{t}$ cross sections in the signal and control region avoids much of the 
  theoretical systematic uncertainties.  
  In Ref. \cite{Kauer:2004fg} the theoretical uncertainty 
  on the ratio $\sigma_{signal\_reg}/\sigma_{control\_reg}$
  has been studied at parton level with LO precision 
  by varying the renormalization and factorization scale. 
  The error has been estimated to range 
  between $3\%$ to $10\%$, mostly due to the choice of the PDF. 
  For what it has been shown before, the theoretical error
  can be larger because of other factors, mainly the parton shower model.  
  A 10\% systematical error due to theoretical uncertainty will be assumed as
  reported in Ref. \cite{Kauer:2004fg}, although baring in 
  mind that this could be an optimistic estimation.

\subsection* {\bf Jet Energy Scale uncertainty.} 

\begin{figure}[htbp]
    \begin{center}
      \includegraphics[scale=0.8]{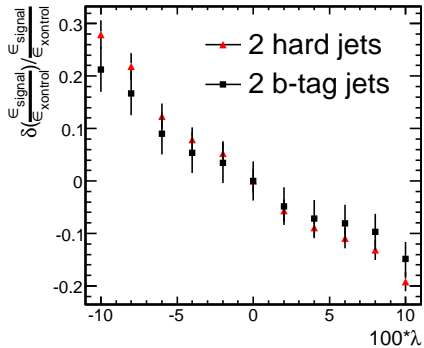}
    \end{center}
    \caption{Relative variation of $\frac{\epsilon_{signal\_reg}}{\epsilon_{control\_reg}}$
      as a function the jet momentum rescaling factor ($\lambda$). The red triangles
      represent the control region defined by two hard jets whereas the black squares
      correspond to the two b-tagged jets phase space area.
    }
  \label{JESUncert}
  \end{figure}

  In the background normalization procedures we proposed, the jet energy scale (JES) 
  uncertainty
  is particularly important since it affects in opposite manners the signal region, 
  defined by vetoing the jets, and the control region where the presence of two jets is 
  required. To take into account this sort of anticorrelation of $\epsilon_{signal\_reg}$ 
  and $\epsilon_{control\_reg}$, we estimate the effect of the JES uncertainty directly
  on their ratio by rescaling the measured jet four momentum by an amount corresponding
  to the percentual uncertainty (i.e. $P^{\mu}_{jet} = (1+\lambda)P^{\mu}_{jet}$). \\
  In the plot of Figure (\ref{JESUncert})  the relative 
  variation of $\frac{\epsilon_{signal\_reg}}{\epsilon_{control\_reg}}$ for various values of 
  $\lambda$
  is shown. In the plot the triangles represent the control region defined by
  requiring two jets with $E_t$ greater then 50 and 35 GeV,
  whereas the squares stand for the control region defined
  by requiring two b-tagged jets\footnote{
    The reason way the ratio $\epsilon_{signal\_reg}/\epsilon_{control\_reg}$ in the latter
    case is less sensitive to the JES uncertainty is that the $E_t$ threshold for
    the b-jets candidates is $20~GeV$ and the fraction of \ttbar events 
    with b-tagged jets with $E_t$ close to that threshold is very small.
  }.
  A realistic estimation of the JES uncertainty at CMS after integrating
  $10~fb^{-1}$ of LHC is ${\cal O} (5\%)$. The the corresponding relative variation
  of $\rm \epsilon_{signal\_reg}/\epsilon_{control\_reg}$ is  
  $\rm \sim8\%$ for the double b-tagging defined control region 
  and $\rm \sim10\%$ for the two high $\rm E_t$ jets control region.

\vspace{-5 mm}

\subsection* { $\alpha$ criterion uncertainty.} 

  To estimate the systematic uncertainty due to $\alpha$ criterion, the value of the cut  
  has been varied from 0.15 to 0.25. Moreover different values of the
  minimum $\rm p_t$ for a track to be included in the sum have been
  tried, from 2 to 3~GeV. The consequent variation of the jet veto 
  efficiency ($\rm \epsilon_{signal\_reg}$) is relatively small, i.e. of the order of 4\%.

\subsection* { b-Tagging uncertainty.} 

  In Ref. \cite{BTagSys:2005cn} the precision with which the b-tagging efficiency 
  will be known at CMS is expected to be $11\%$ for $1~fb^{-1}$ integrated luminosity
  and it is foreseen to improve till $7\%$ with $10~fb^{-1}$. These values
  represent directly the uncertainty on $\epsilon_{control\_reg}$ in the case
  of the control region defined by requiring two b-tagged jets.

\subsection* { Uncertainties on $N_{control\_reg}$.} 

  It has been shown in the previous section that \ttbar is plainly the 
  dominant process in both the proposed control regions. 
  In the worst case, i.e. when the control region
  is defined by two high $E_t$ jets, the fraction of events coming form
  other processes is smaller than 4\%.
  Provided that this fraction is small, it is safe to simply 
  neglect this source of systematic.

\vspace{5 mm}

For $10~fb^{-1}$ the experimental uncertainties listed above accounts for a systematic
error of $\sim{11}\%$ for both the \ttbar control regions. Including
the theoretical uncertainty this error does not exceed $16\%$.

\subsection*{Statistical uncertainties}

The statistical precision with which the number of \ttbar events in the 
signal region can be known depends on the expected number of \ttbar 
events in the control region. From the numbers quoted in Table~\ref{doublebTag_table},
assuming a poissonian behavior, it is clear that 
the error due to systematic uncertainties is predominant with
respect to the statistical ones for both the proposed normalization procedures.

\section{Conclusions}

The searches for the Higgs boson at the LHC via the channel \signal offers the possibility
to study the \ttbar and $Wt$ in a peculiar phase space region. This represents
a major challenge either from the theoretical and from the experimental point of 
view. \\ 
The present status of the Monte Carlo tools for \ttbar simulation has
been discussed. 
The comparison between a set of generators shows that
the NLO effects can be safely accounted as a global rescaling of the
total cross section. On the contrary the PYTHIA and HERWIG showering models 
differ in predicting jets multiplicity and energy spectra. 
Finally the spin correlation between the top-pair induces a variation of ${\cal O}(10\%)$
in the leptons' selection cuts efficiency.
In section \ref{tWDescription}
it has been show that the  
$Wt$ process can be reliably calculated at NLO in the signal region defined
by a jet veto without double counting with \ttbar. \\
Finally the normalization of the \ttbar background have been discussed. 
Two control regions have been proposed, one based on b-tagging the 
jets coming from the top-pair and the other by requiring two high \ET jets.
Both approaches provide a reliable phase space area dominated by \ttbar events
and lead to an overall systematic uncertainty of 16\%.


\begin{thebibliography}{9}

\bibitem {dittmar} {
  M.~Dittmar and H.~K.~Dreiner,
  \emph{How to find a Higgs boson with a mass between 155-GeV to 180-GeV at the
  LHC},
  Phys.\ Rev.\ D {\bf 55} (1997) 167
  [arXiv:hep-ph/9608317].
}

\bibitem{ASGD} { G.~Davatz, M.~Dittmar, A.S.~Giolo~Nicollerat, \emph{Standard Model 
  Higgs Discovery Potential of CMS in the $\rm H\to
  WW\to \ell\nu\ell\nu$ Channel}, {\bf CMS NOTE-2006-048} . 
}

\bibitem{herwig} {
  G.~Corcella {\it et al.}, \emph{
  HERWIG 6.5 release note}, [arXiv:hep-ph/0210213].
}

\bibitem{mcatnlo} {
  S.~Frixione and B.~R.~Webber, \emph{
    Matching NLO QCD computations and parton shower simulations},
  JHEP {\bf 0206} (2002) 029
  [arXiv:hep-ph/0204244] \\
  S.~Frixione, P.~Nason and B.~R.~Webber,
  \emph{Matching NLO QCD and parton showers in heavy flavour production},
  JHEP {\bf 0308} (2003) 007
  [arXiv:hep-ph/0305252].
}

\bibitem{ttCMSPaper} { G.~Davatz, A.S.~Giolo~Nicollerat, M. Zanetti,
  \emph{Systematic uncertainty of the top background in the  $\rm H\to
    WW\to $ Channel}, {\bf CMS NOTE-2006-048}. 
}

\bibitem{pythia} T. Sj\"ostrand et al., Comput. Phys. Commun. 135 (2001) 238.

\bibitem{TopREX} S.R. Slabospitsky, Comput. Phys. Commun. 148 (2002) 87.

\bibitem{maltoniCoimbra} F. Maltoni, in Proceedings of TOP2006 workshop.

\bibitem{campbell}{
  J.~Campbell and F.~Tramontano,
  \emph{Next-to-leading order corrections to W t production and decay},
  [arXiv:hep-ph/0506289].
}

\bibitem{mcfm}{
  J.~Campbell and K.~Ellis \emph{Monte Carlo for FeMtobarn processes},
http://mcfm.fnal.gov/
}

\bibitem{BTagging:2005an} {
  {\bf  CMS NOTE-2006/019 }, A. Rizzi, F.~Palla and G. Segneri, {\em Track impact parameter based b-tagging with CMS}
}

\bibitem{Maltoni:2002qb}{
  F.~Maltoni and T.~Stelzer,
  JHEP {\bf 0302} (2003) 027
  [arXiv:hep-ph/0208156].
}

\bibitem{Kauer:2004fg} {
  N.~Kauer \emph{Top Background Extrapolation for $\rm H \ra WW$ Searches at the LHC}, 
  Phys.\ Rev.\ D {\bf 70} (2004) 014020
  [arXiv:hep-ph/0404045].
}

\bibitem{BTagSys:2005cn} J. Heyninck, in Proceedings of TOP2006 workshop.




\end{thebibliography}
\end{document}